\documentclass[preprint,preprintnumbers,amsmath,amssymb]{revtex4}

\usepackage{amsmath}
\usepackage{amssymb}
\usepackage{bm}
\usepackage{indentfirst}
\usepackage{mathtools}
\usepackage{braket}
\usepackage{amsthm}
\usepackage{multirow}
\usepackage{threeparttable}
\usepackage{setspace}
\usepackage{tikz}
\usepackage{verbatim}
\usepackage{graphicx}

\begin{document}
\title{Greenberger-Horne-Zeilinger test for multi-dimension and arbitrary time nodes entangled histories}
\author{Junkai Dong}
\affiliation{$^1$ Department of Physics, Cornell University, New York 14853-2501, USA \\ $^2$ Fuzhou No.1 High School, Fuzhou 350108, China}

\author{Yi-Ming Chen}
\affiliation{$^3$ Department of Physics, Tsinghua University, Beijing 100871,  China}

\author{Da Xu}\email{xuda2016@pku.edu.cn}
\affiliation{$^4$ State Key Laboratory for Mesoscopic Physics and School of Physics, Peking University, Beijing 100871,  China}

\author{Zhang-Qi Yin}\email{yinzhangqi@mail.tsinghua.edu.cn}
\affiliation{$^5$ Center for Quantum Information, Institute for Interdisciplinary Information Sciences, Tsinghua University,
Beijing 100084, China}

\begin{abstract}

Based on the framework of consistent history theory, the quantum entangled history was proposed in 2015 and experimentally verified through temporal Greenberger-Horne-Zeilinger (GHZ) test with $3$ time nodes in 2016.
In this paper, we extend the temporal GHZ test to arbitrary time nodes and even system dimensions. Then, we define a witness to distinguish between the quantum entangled histories and the classical histories.
The minimums of the witness for the classical histories are calculated for arbitrary number of time nodes and the system dimensions $2$ and $\infty$. It is found that the minimums of the witness for the classical histories is always larger than the quantum entangled histories minimum $-1$. Only when both the number of time nodes and system dimensions approach to infinity, the minimum of the witness for classical and quantum entangled histories are identical.


Received: 05-Jul-2017; Revised: 06-Aug-2017; Accepted: 08-Aug-2017;

\end{abstract}

\maketitle

\section{Introduction}

\setcounter{page}{1}
Quantum entanglement, since proposed by  Einstein, Podolsky and Rosen (EPR) \cite{EPR} and further explored by Schr\"odinger \cite{sch} in 1935, has always been the focus of quantum physics realm. The EPR paradox revealed the conflict between quantum theory and local realism.
Almost 30 years later, in 1964, John Bell first came up with the prototype of a family of inequalities, which were later called Bell inequality \cite{Bell1,Bell3,CHSH}, to express certain limitation that every local classical hidden variable theory should follow up. Therefore, it could be used to distinguish the quantum theory from the local hidden variable theories
\cite{Guo}. Experimental verifications on Bell inequlity lasted for 40 years, until the loophole-free experiments was performed in 2015 \cite{exp1, exp2, exp3}.\par
The widely accepted interpretation of quantum mechanics is the Copenhagen interpretation. However, one of the major problems with the interpretation is the unnatural collapsing of states when a quantum state is measured. Due to this concern, Griffiths brought up a different interpretation, which can give the same physical result as Copenhagen interpretation but without collapsing of states, called the consistent histories theory \cite{R.Griffiths}. Under the framework of consistent histories theory, Frank Wilczek and Jordan Cotler defined
a new concept quantum entangled histories \cite{Wilczek1}, which are entanglement in time, other than entanglement in space. Later, they proposed a Bell test for entangled histories \cite{Wilczek2}.
We should note that some previous literature studied temporal entanglement both theoretically \cite{Brukner,Fritz,Paz,Emary} and experimentally \cite{Waldherr}. They focused on the paradox emerging from entanglement induced by measurement and prediction by classical theory. However, the entangled history theory focuses on the intrinsic correlation in quantum dynamics.

In 2016, the quantum entangled history was experimentally verified through a temporal Greenberger-Horne-Zeilinger (GHZ) test \cite{GHZ,Mermin} for quantum entangled history state
with $3$ time nodes \cite{Yin}. The classical stochastic processes were introduced as the representative of classical theories. A function $G$ was defined to distinguish quantum and classical theory. It was proved that for quantum theory, $G$ could approach $-1$ while the lower bound of $G$ for classical theory is $-\frac{1}{16}$. In the experiment, $G$ was measured of $-0.656$, which clearly showed that quantum entangled histories existed.
\par
This paper aims to broaden the scope of temporal GHZ paradox from $3$ time nodes to arbitrary nodes, and from dimension $2$ (qubit) to arbitrary even (qudit).
For the $2$ dimensional system, we discuss a temporal GHZ-type test with arbitrary time nodes. We define a witness and prove that the boundaries between classical and
quantum entangled histories expectations exist. We find exact boundary formula for arbitrary time nodes $m$.
Inspired by Ref. \cite{Tang}, we construct the temporal GHZ-type test for high dimensions (qudit). The boundaries between classical and quantum expectations
are also proved to be existed and calculated. We specifically analyze the behavior of minimum when the dimension is $2$ and $\infty$. We find that when the dimension and number of time nodes tend to infinity, the minimum will be approached to $-1$. Therefore, the classical and quantum predictions are indistinguishable.\par

This paper is organized as follows. Section II focuses on the background knowledge and mathematical framework of entangled history. Section III gives a brief review on the GHZ type tests in
space. Section IV discuss the temporal GHZ-type tests. The boundaries between classical and quantum entangled histories predictions are calculated and proved.
In the last section, we give a brief summary and prospect.

\section{Theoretical Basis for Entangled Histories}
The introduction of main mathematical formulation of entangled history theory mainly follows the structure of \cite{Wilczek1}, where the motivation of entangled history theory is discussed more in detail.
The Hilbert space of history states is the vector space which we will focus on. It is defined as the tensor product of several ordinary Hilbert spaces, each simply the Hilbert space of the system at a particular time $t_{i}$. An issue worthy of attention is that the time sequence is from later to former, i.e., the history Hilbert space should be written as follows \cite{Wilczek1,R.Griffiths}:
\begin{equation}
\check{\mathcal{H}} \coloneqq \mathcal{H}_{t_n} \odot \mathcal{H}_{t_{n-1}} \odot \dots \odot \mathcal{H}_{t_1}, \qquad t_n>t_{n-1}>\dots>t_1
\end{equation}
in which the special notation $\odot$ is used to represent tensor product in time domain as in  \cite{Wilczek2} and reserve the notation of $\otimes$ to represent tensor product in space domain.
\par In this paper, the Hilbert space of the history of a sequence of discrete moments, each connected by a bridging operator, is concerned. The bridging operator is denoted $T(t_j,t_i)$ for mapping the Hilbert space $\mathcal{H}_{t_i}$ to $\mathcal{H}_{t_j}$, and is determined using the Schr\"odinger's Equation.
The history states are defined as:
\begin{equation}
|\Psi)=P^{i_n}_{t_n}\odot\dots\odot P^{i_1}_{t_1}
\end{equation}
in which $P^{i_k}_{t_k}$ is some projector in $H_{t_k}.$ Each $t_k$ is called a time node.\par
Now consider a GHZ history state
\begin{equation}\label{eq:tGHZ}
|GHZ)=\frac{1}{\sqrt{2}}([0]\odot[0]\odot[0]-[1]\odot[1]\odot[1])
\end{equation}
in which $[i]=\ket{i}\bra{i}$. An important characteristic of the measurement of history states is that they must be constructed and measured spontaneously. An example may be the measurement of the GHZ history state shown in Ref. \cite{Cotler} that includes the protocol for measuring history states. Using the formalism of Ref. \cite{Cotler}, we can find the expectation of a temporal observable, $Q$, in the same way we calculate the expectation of a normal observable $Q'$, namely $\bra{\psi}Q'\ket{\psi}$. The expectation of the temporal observable is $\bra{i_1i_2\dots i_n}Q\ket{i_1i_2\dots i_n}$, in which $[i_1i_2\dots i_n]=P^{i_n}_{t_n}\odot\dots\odot P^{i_1}_{t_1}$.
If a GHZ state $\ket{GHZ}=\frac{1}{\sqrt{2}}(\ket{000}-\ket{111})$ is constructed and measured in the $\ket{000},\ket{001}\dots$ basis, probability amplitudes $\braket{ijk|GHZ}=\frac{1}{\sqrt{2}}(\braket{i|0}\braket{j|0}\braket{k|0}-\braket{i|1}\braket{j|1}\braket{k|1})$ are obtained. In experiment, the measurement needs auxiliary qubits or qudits to record the information of the system.

The probability of some measurement outcome from a history state is identical to the probability of measuring a normal state and get the same results, namely, the probability of getting outcomes $(i,j,k)$ is also $\braket{ijk|GHZ}$. Due to this property, whenever calculation of the expectation for a history state is needed, we use the inner product of bra and ket as usual.

\section{GHZ-type entanglement in space}
The GHZ-type entanglement is one of the most well-studied type of entanglement since it demonstrates distinctive results predicted by classical local theories and quantum theories
\cite{GHZ,Mermin}. In this section, the current results and construction of several others are summarized. These examples in space domain will provide significant support and a general framework to our discussion about the GHZ-type tests in time domain. From this section, we omit any notation of tensor product in time.
\subsection{Original GHZ construction}
The original GHZ state\cite{GHZ,Mermin} is a three-partite two-dimensional entangled state:
\begin{equation}
\ket{GHZ}=\frac{1}{\sqrt{2}}(\ket{000}-\ket{111})
\end{equation}
Witnesses denoted $Q_1={X_1}{X_2}{X_3},\ Q_2={X_1}{Y_2}{Y_3},\ Q_3={Y_1}{X_2}{Y_3},\ Q_4={Y_1}{Y_2}{X_3}$ are used, where $X_i$ or $Y_i$ is the pauli matrix $X$ in the $i$th Hilbert space.
\begin{equation}
\begin{aligned}
&\braket{{X_1}{X_2}{X_3}}=-1,\ \braket{{X_1}{Y_2}{Y_3}}=1,\ \braket{{Y_1}{X_2}{Y_3}}=1,\ \braket{{Y_1}{Y_2}{X_3}}=1.
\end{aligned}
\end{equation}
$\ket{GHZ}$ is a common eigenvector of all four operators. An observable $G={X_1}{X_2}{X_3}{X_1}{Y_2}{Y_3}{Y_1}{X_2}{Y_3}{Y_1}{Y_2}{X_3}$ is measured. Hence, $G_{qm}=\braket{{X_1}{X_2}{X_3}}\braket{{X_1}{Y_2}{Y_3}}\braket{{Y_1}{X_2}{Y_3}}\braket{{Y_1}{Y_2}{X_3}}=-1$ in quantum theory. As $\ket{GHZ}$ is a common eigenvector, $G_{qm}=\braket{{X_1}{X_2}{X_3}{X_1}{Y_2}{Y_3}{Y_1}{X_2}{Y_3}{Y_1}{Y_2}{X_3}}$. If $G$ is considered in classical local theory, the incommutativity of the operators is lost, and thus $G_{c}=\braket{\prod{Q_i}}=(X_1 X_2X_3Y_1Y_2Y_3)^2=1$ because each operator is treated like a random variable with value $\pm1$. This is a distinctive difference.\par
An important advantage of GHZ-type entanglement is that the prediction of quantum mechanics and classical stochastic theory is determined and separated. Hence, it is easier for the experiments to detect GHZ-type entanglement.
\subsection{Extension to higher dimension and arbitrary number of particles}
When the GHZ-type entanglement is extended to higher dimensions, we aim to preserve the advantages of GHZ paradox: the quantum prediction and the classical prediction are significantly separated from each other and the witnesses are all products of $X$, $Y$ and $Z$, the generators of the Heisenberg group.
The operators $X$, $Y$ and $Z$ are defined as follows
\begin{equation}
\begin{aligned}
&X=\sum_{k=1}^{d-1} \ket{(k+1)\mod d} \bra{k}\\
&Y=\sum_{k=1}^{d-1} e^{2\pi\mathrm{i}k/d} \ket{(k-1)\mod d} \bra{k}\\
&Z=\sum_{k=1}^{d-1} e^{2\pi\mathrm{i}k/d} \ket{k} \bra{k}\\
\end{aligned}
\end{equation}

Previously, the genuine GHZ paradoxes are constructed for even dimensions and arbitrary number of particles  \cite{Tang,Zhang,Su2017}. They constructed special graphs called GHZ graphs whose adjacency matrix and vertex operators give rise to a GHZ-type paradox. This study provides us with an ideal model of entangled histories.\par
We have found no construction of an odd dimension GHZ paradox using the same definition as ours in previous literature. In these papers \cite{Cerf,Tang}, the construction is only given for even dimension. A proof that there is no GHZ paradox in the framework of odd dimension is given in Appendix A. However, if we use another definition of operators, the GHZ paradoxes in odd dimension can be defined, as shown in Ref. \cite{Ryu1,Ryu2,Lawrence}. However, their definition needs special calculation for each pair of particle number and dimension in order to control the phases of eigenvalues to reach a paradox. The construction for the GHZ paradoxes in odd dimensions is state-dependent.

\section{Temporal GHZ tests with arbitrary time nodes and dimensions}
A complete construction of entanglement witnesses for GHZ states in space has been summarized in the last section.  In this section, we explore GHZ-type entangled histories for arbitrary time nodes and dimensions. We construct the GHZ-type tests for entangled history states. Similar as Ref. \cite{Yin},
we find that there are boundaries between entangled histories and the classical histories. \par
Similarly as GHZ test in space, we can define an observable $G$ to distinguish quantum entangled histories and classical states.
The quantum prediction of $G$ for entangled GHZ-type history state, e.g. Eq. \eqref{eq:tGHZ} is always $-1$. In classical theory, each time nodes in histories are correlated in a non-local way, rather than locally related in GHZ states in space. Hence, instead of taking $\braket{\prod{Q_i}}$ for classical mechanics, the observable $\prod{\braket{Q_i}}$ is taken to signify the reduced reliability of $Q_i$ on each other. Note that $\prod{Q_i}$ is still $1$.\par
Hence, each possible combination of values of $Q_i$ - a timeline - is taken to be $a_j=(Q_{ij})$, in which $Q_{ij}$ is the $i$th outcome of the combination $a_j$. Suppose the probability for $a_j$ is $p_j$. Then the quantity $\prod{\braket{Q_i}}$ can be expressed as:
\begin{equation}
E_t(n,d)=\prod_i(\sum_j Q_{ij} p_j)
\end{equation}
in which $n$ is the number of witnesses and $d$ is the dimension of the Hilbert space.
Now, the problem reduces to finding the boundary for $E_t(n,d)$. Also, we denote the number of time nodes $m$. In general, $n=m+1$. Hence, $n$ grows when $m$ increases.
\subsection{Temporal GHZ test for Qubits}

In Ref. \cite{Yin}, the minimum of $E_t(4,2)$ was calculated and proved. This corresponded to a qubit system with $3$ time nodes. In the paper, they proved that $E_t(4,2)$ has a minimum of $-\dfrac{1}{16}$. However, the method in Ref. \cite{Yin} cannot easily extend to arbitrary time nodes $m\geq 3$.\par

Here we consider an entangled GHZ-type history state with number of time nodes $m\geq 3$. It is easily found that here the number of witnesses $n=m+1$.
In this formalism, there would be $2^n$ different history timelines with outcome $1$ or $-1$ for the $n$ measurements, or witnesses. One very crucial issue is that if we multiply all the outcomes of a timeline, the result should be $1$. In mathematical form, it is:
\begin{equation}
\prod_{i}{Q_{ij}}=1
\end{equation}
Because changing the last outcome from $1$ to $-1$ or $-1$ to $1$ changes the sign of the product, it can be concluded that there are $2^{n-1}$ possible outcomes.
\par Suppose outcome $j$ has a probability $p_j$ assigned to it. Then the classical expectation in time domain, can be expressed as
\begin{equation}
E_t=\prod_{i=1}^{n}{\left(\sum_{j=1}^{2^{n-1}}{{Q_{ij}}p_j}\right)}
\end{equation}
This is a polynomial for $p_j$ with the constraint that $\sum_{j}{p_j}=1$.
\par We have to find the minimum for $E_t$ to confirm that it is indeed seperated from quantum outcomes. In fact, the ultimate result is
\begin{equation}E_t(n,d)\in[-(1-\frac{2}{n})^n, 1].\end{equation}
The detailed calculation can be found in Appendix B.\par
The importance of the minimum lies in two aspects. First, surprisingly, the minimum is not reached in a maximally mixed timeline, in which each of the timeline has the same probability. Furthermore, the combination which generates the minimum is unsymmetrical. Second, as shown in Fig. \ref{fig:boundary}, the lower bound is not $-1$ when $n\rightarrow +\infty$. In fact, $\lim_{n\rightarrow +\infty}{E_t(n,2)}_{min}=-\text{e}^{-2}$, which is larger than $-1$. Hence, a gap is observed between the quantum prediction and classical prediction.
For $n=4, d=2$, the GHZ-type test for entangled histories was performed with single photon experiment \cite{Yin}. The quantum and classical predictions gap we proved here makes the GHZ-type entangled histories tests for arbitrary time nodes possible in experiment.
\subsection{Estimations for higher dimensions}
In higher dimensions, by the construction of witnesses, $n=m+1$. $Q_{ij}$ takes the positive powers of $\exp{(2 \pi i/d)}$. $E_t$ should be real while each sum in $j$ may not be real. This generates a substantial problem for calculating $E_t$ for $d\leq4$ since there is no clear and feasible way to calculate the argument of $E_t$. Furthermore, since $\epsilon^k$ is discrete on the unit circle, we cannot use analytic methods if $d\neq\infty$. These are the main difficulties in calculating. \par
\begin{figure}
\includegraphics[width=0.6\textwidth]{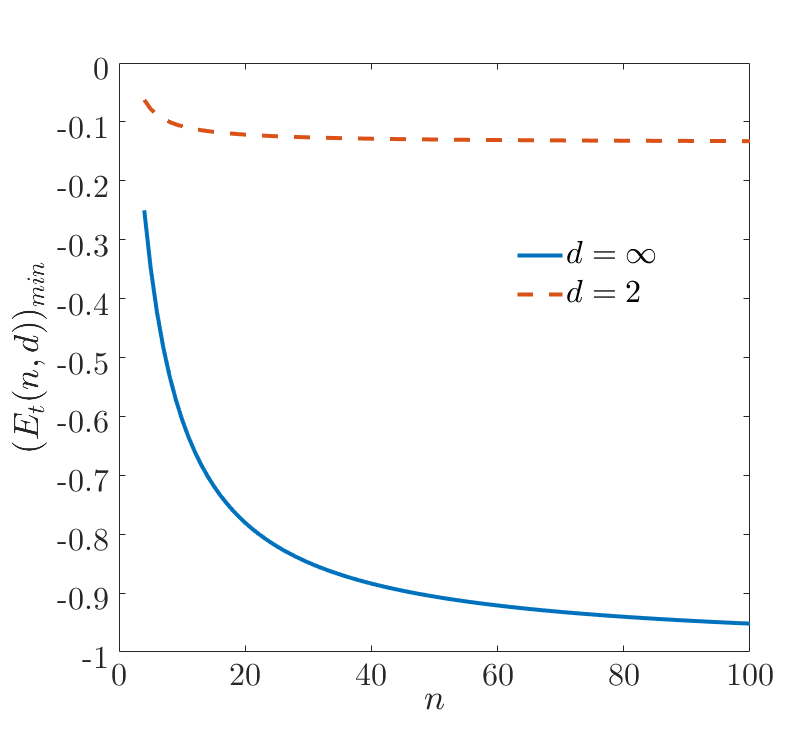}

\caption{The boundaries $E_t(n,2)$ ($E_t(n,\infty)$) between GHZ-type entangled histories and classical histories for Hilbert space dimension $2$ ($\infty$) and witness number $n$. $E_t(n,\infty)$ will approach $-1$ when $n$ approaches $\infty$. }\label{fig:boundary}
\end{figure}

However, the minimum of $E_t(n,\infty)$ can be calculated. Since the phase could be set as continuous  when $n\rightarrow \infty$, the optimization is possible. The main idea of calculation is to find the deviation of the phase between entries of the timelines and the ultimate expectations of the witnesses.The deviations conform to some restraints, as shown in Appendix C,  we find out that under the restraint the minimum is $-(\cos{\pi/n})^n$. Also, the construction of the situation which generates the minimum requires that $n$ can divide $d$. Hence the minimum is reached for infinite times for fixed $n$ when we increase $d$. There will be a fluctuating pattern, while the deviation gradually decreases when $d$ is increased.
The Fig. \ref{fig:boundary} shows the minimum of $E_t(n,\infty)$ with respect to the number of witnesses $n$. It is found that, the boundaries for $d$ approaching to $\infty$ is
much lower than the boundaries of $d=2$ for every $n$. Besides, we can see that when $n\to\infty$, the minimum of $E_t(n,\infty)$ becomes $-1$.
In other words, the quantum and classical predictions are mixed under this condition.
\section{Conclusion and prospect}
In this paper, we analyzed the GHZ-type entangled histories for arbitrary time nodes and dimensions. In particular, the case of $d=2$ and $d=\infty$, are explored. We introduced classical correlations in time which give rise to an observable called $E_t(n,d)$. We prove respectively that the minimum of $E_t$ for $d=2$ and $d=\infty$ are $-(1-\frac{2}{n})^n$ and $-(\cos{\frac{\pi}{n}})^n$. They are both larger than the quantum prediction  $-1$ for finite number of time nodes $m=n-1$. \par
Moreover, there is an interesting phenomenon. Usually if we increase the dimension of Hilbert space $d$ to infinity, the quantum system would tend to behave in a  classical way. However, in GHZ-type tests for entangled histories, even if $d=\infty$, there is still a huge gap between classical and quantum predictions for finite $m$. Only if we increase both $d$ and $m$ (with $n$) to infinity, which means both system dimensions and time are continuous, the predictions of both quantum and classical theories are indistinguishable. Though there is no dissipating channel being introduced, the mixture of quantum and classical predictions is simultaneous.

This phenomenon means that when $d$ is infinite, though the quantum system is similar to a complex classical system, there are still fundamental differences between quantum and classical correlation. For small $n$, if we observed a measurement outcome lower than the bound given, we can conclude device-independently that there is indeed quantum entanglement, even in time. We have not proved the minimum of $E_t(n,d)$ for all combinations of $n$ and $d$. Further calculation will help us understand how the dimension of the system and the number of time nodes change the boundaries between quantum entangled and classical histories. Besides, it may reveal the deep quantum correlation patterns between space and time.

In order to experimentally test the theory of the present work, beside the single photon experiments \cite{GHZ}, we may use the NMR quantum simulator \cite{Jin}, the trapped ions \cite{An2015}, the graphene \cite{Sil}, or the optically trapped nano-particles \cite{Yin2013,Zych}.
This work may stimulate further studies. For example, in future we may investigate the entangled histories for living object \cite{Li}, experimentally testing the genuine entangled histories without sharing references \cite{Wang}, etc.

\section*{Acknowledgments}
We would like to thank Tongcang Li for valuable suggestions and insights. We would also like to thank Dr. Ryu for his special insights for the discussion of odd dimension.
This work is funded by the National Natural Science Foundation of China NO. 61435007, and
the Joint Fund of the Ministry of Education of China (6141A02011604).
\section*{Appendix A: No GHZ paradox in odd dimentions}

First, it can be observed that in the GHZ paradox, the quantum expectation has to be $-1$ because the $d$th power of some random variable is used to generate the certain result of $1$ in classical mechanics. As the GHZ state constructed has to be an eigenvector of the witnesses with real eigenvalues, one of the eigenvalues must be $-1$. The reason is that $X$, $Y$ and $Z$ are all unitary operators, thus their tensor product must be unitary, and unitary operators have eigenvalues with module $1$.\par
Thus, it remains to show that with the operators defined above, we can not generate any eigenvalue of $-1$.\par
$X$, $Y$ and $Z$ all have the spectrum of $S=\{\epsilon^n|n\in\mathbb{Z}\}$ with $\epsilon=e^{2\pi\mathrm{i}/d}$. On each sub-Hilbert space $X$, $Y$ or $Z$ or their arbitrary product is applied. A basis of the eigenspaces is taken to form the basis for the qudit with eigenvalue lying in $S$. Some of the eigenvalues might degenerate. Taking tensor product for each sub-Hilbert space, a basis for the entire Hilbert space is formed. Dividing it into the eigenvectors of the complete tensor product of $X$, $Y$ and $Z$ defined on each sub-Hilbert space, the eigenvalues still lie in $S$ as arbitrary products of the operators have order $d$.\par
However, as $d$ is odd, $-1\notin S$. This completes the proof.\par
If we consider the definition of operators in these papers \cite{Ryu1,Ryu2,Lawrence}, we observe that the eigenvalues of the operators are not in $S$ defined above. The GHZ paradoxes
can be constructed under this stated dependent method.
\section*{Appendix B: GHZ Test for quibits with arbitrary number of time nodes}

\par We want to prove the conjecture that
\begin{equation}
(E_t)_{min}=-\left(\frac{n-2}{n}\right)^n
\end{equation}
\par The proof is as follows:
\begin{proof}
\par First, a solution is given to generated the desired outcome:
\begin{equation}
\begin{aligned}
&\ 1,1,1,\dots,1\\
&-1,-1,1,\dots,1\\
&-1,1,-1,\dots,1\\
&\dots\\
&-1,1,1,\dots-1\\
\end{aligned}
\end{equation}
\par Choose these $p_j$ to be $\frac{1}{n}$ and others to be $0$, this situation yields the value $-(\dfrac{n-2}{n})^n$.
\par Change the sign of $Q_{ij}$ when $i=1$ and obtain $Q'_{ij}$. The corresponding $E'_t$ is
\begin{equation}
E_t^{'}=\prod_{i=1}^{n}{\left(\sum_{j=1}^{2^{n-1}}{{Q'_{ij}}p_j}\right)}
\end{equation}
also
\begin{equation}
(E_t)_{min}=-(E_t^{'})_{max}
\end{equation}
\par We use the Arithmetic-Geometric Average Inequality to get
\begin{equation}
(E_t^{'})\leq\left(\frac{\sum_{i=1}^{2^{n-1}}d_j p_j}{n}\right)^{n}
\end{equation}
in which
\begin{equation}
d_j=\sum_{j=1}^{m}{Q'_{ij}}
\end{equation}
\par More attention should be paid here in order to demonstrate that the inequality can be used. The inequality demands that every term  $\{\sum_{j=1}^{2^{n-1}}{{Q'_{ij}}p_j}\}$ must be larger than or equal to $0$, which is not necessarily the case here. However there is a simple argument that helps us get out of this. If the product is negative, the inequality fails, but it is obviously less than $(\frac{n-2}{n})^n$ and this situation should be ignored in search for the maximum. If the product is positive, there must be an even number of negative signs. $-1$ is multiplied on each of the previously negative sums. The whole product is the same.
\par But this time,
\begin{equation}
|d_j|_{max}=n-2
\end{equation}
since
\begin{equation}
|d_j|_{max}\leq n
\end{equation}
but for the equality to hold, all $Q'_{ij}=1$ or $Q'_{ij}=-1$, but $Q'_{ij}$ cannot be all the same since $\prod_j Q'_{ij}=-\prod_j Q_{ij}=-1$. The maximum is not reachable. $|d_j|$ is an even number because $n$ is even. However, when we choose $Q'_{i1}=(-1, 1, 1, \dots,1)$
\begin{equation}
|d_1|=n-2
\end{equation}
which is the largest even number less than $n$. So the maximum is proven.\par
Thus
\begin{equation}
\left(\frac{\sum_{i=1}^{2^{n-1}}d_j p_j}{n}\right)^{n}\leq\left(\frac{|d_j|_{max}}{n}\right)^n=\left(\frac{n-2}{n}\right)^n
\end{equation}
\par So
\begin{equation}
(E_t)_{min}=-(E_t^{'})_{max}=-\left(\frac{n-2}{n}\right)^n
\end{equation}
\par Similarly, the situation of taking equalities in the inequalities is verified, and the solution constructed meets all the standards.
\end{proof}
\par The minimal value reached by increasing $n$ to infinity would be
\begin{equation}
\lim_{n \rightarrow +\infty}-\left(\frac{n-2}{n}\right)^n=-\mathrm{e}^{-2}.
\end{equation}
\section*{Appendix C: Minimum for $E_t(n,\infty)$ and $E_t(n,kn)$}
\begin{proof}
Since $d=\infty$, $Q_{ij}$ can be any complex number on the unit circle. Suppose $p_j$ is $0$ for all but finite $j$ to simplify the situation.  We want $E_t(n, \infty)$ to be as negative as possible; to do that, consider each term of $E_t$. Take a combination of $Q_{ij}$ and $p_j$ which satisfy the restraints. Suppose $a_i=\sum_j Q_{ij} p_j=r_i\exp{(i \alpha_i)}$. To make $E_t$ negative it is required that $\sum_i\alpha_i = \pi +2k\pi, k\in\mathbb{Z}$. Now suppose $Q_{ij}=\exp{(i \alpha_{ij})}$ and $\epsilon_{ij}\mod \alpha_{ij}-\alpha_i$. Now the restraint that $\prod_i{Q_{ij}}=1$ becomes $\sum_i \epsilon_{ij} = \pi +2k\pi, k\in\mathbb{Z}$. Also $\epsilon_{ij} \in [-\pi,\pi]$. The expression for $E_t$ becomes
\begin{equation}
E_t=-\prod_i(\sum_j \cos{\epsilon_{ij}} p_j)
\end{equation}
Now $E'_t=\prod_i(\sum_j \cos{\epsilon_{ij}} p_j)=\prod_i b_i$ should be considered; it should be as large as possible. Eliminate all the possibilities that $E'_t$ are negative, similar to the process in Appendix B, since the minimum of $E'_t$ - which correspond to the maximum of $E_t$ - is $-1$. If $E'_t$ is positive, multiply each negative $b_i$ by $-1$ which would not change the value of $E'_t$ and still preserving the restraints. Now that each $b_i$ of $E'_t$ is positive, the arithmetic-geometric inequality can be applied:
\begin{equation}
E'_t=\prod_i(\sum_j \cos{\epsilon_{ij}} p_j)\leq(\frac{\sum_j p_j\sum_i\cos{\epsilon_{ij}}}{n})^n\leq (|\sum_i\cos{\epsilon_{ij}}|_{max}/n)^n
\end{equation}
which increases when $|\sum_i\cos{\epsilon_{ij}}|$ increases. Now the Lagrangian multiplier method with restraint is applied. This is the most significant difference of the case $d=\infty$ with the cases $d<\infty$. The choice of $Q_{ij}$ is continuous, rather than discrete. When $d$ is finite, the choice of possible timelines must yield to a transcendental equation, which greatly increases the difficulty of the problem. However, in the case of $d=\infty$, it can be reached that $|\sum_i\cos{\epsilon_{ij}}|\leq n(\cos{\dfrac{\pi}{n}})$ when $n$ is even. When $n$ is odd, the maximum is $n$ but this requires $\epsilon_{ij}=\pi$ which is not possible in physics. The second maximum is $n(\cos{\dfrac{\pi}{n}})$. Hence the conclusion is
\begin{equation}
E_t(n,\infty)\geq -(\cos{\frac{\pi}{n}})^n
\end{equation}
\end{proof}
It is reached when $p_1 = p_2 = \dfrac{1}{2}$, $Q_{1j}=1$, $Q_{2j}=\exp{(2\pi i/n)}$.
Also, when n goes to infinity, $E_t(\infty, \infty)=-1$, which confirms our result.\par

\par
Notice that the same minimum can be reached when $d=kn$ with $k$ a positive integer. We currently do not know about the behavior when $d\neq kn$ but a good guess would be that the minimum is reached when the solution is closest to the desired situation of minimum.\par


\begin{thebibliography}{99}
\bibitem{EPR} A. Einstein, B. Podolsky and N. Rosen, Can Quantum-Mechanical Description of Physical Reality Be Considered Complete? Phys. Rev. {\bf 47}, 777 (1935).

\bibitem{sch}
E. Schr\"odinger, Discussion?of?Probability?Relations?between?Separated?Systems. Mathematical Proceedings of the Cambridge Philosophical Society. {\bf 31} (4): 555¨C563 (1935).

\bibitem{Bell1}
J. S. Bell, On the Einstein Podolski Rosen Paradox. Physics. {\bf 1}, 195 (1964).
\bibitem{Bell3}
J. S. Bell, Speakable and Unspeakable in Quantum Mechanics, 2nd edition. Cambridge, Cambridge UP (2004).

\bibitem{CHSH}
J. F. Clauser, M. A. Horne, A. Shimony and R. A. Holt, Proposed Experiment to Test Local Hidden-Variable Theories. Phys. Rev. Lett. {\bf 23}, 880 (1969).

\bibitem{Guo}
 W J Guo, D H Fan, L F Wei, Experimentally testing Bell's theorem based on Hardy's nonlocal ladder proofs[J]. Science China Physics, Mechanics \& Astronomy, {\bf 58}, 1-5 (2015).


\bibitem{exp1}
M. Giustina, M. A. M. Versteegh, S. Wengerowsky, J. Handsteiner, A. Hochrainer, K. Phelan and Anton Zeilinger, Significant-Loophole-Free Test of Bell¡¯s Theorem with Entangled Photons. Phys. Rev. Lett. {\bf 115}, 250401 (2015).
\bibitem{exp2}
L.K. Shalm, et al. Strong Loophole-Free Test of Local Realism. Phys. Rev. Lett. {\bf 115}, 250402 (2015).
\bibitem{exp3}
B. Hensen, H. Bernien, A. E. Dre'au, Reiserer A., N. Kalb, M. S. Blok and R. Hanson, Loophole-free Bell inequality violation using electron spins separated by 1.3 kilometres. Nature {\bf 526}, 682 (2015).

\bibitem{R.Griffiths}
R. B. Griffiths, Consistent Quantum Theory. Cambridge, Cambridge UP (2002).

\bibitem{Wilczek1}
J. Cotler and F. Wilczek, Entangled Histories. Physica Scripta {\bf T168}, 014004 (2016). arXiv: 1502.02480 (2016).

\bibitem{Wilczek2}
J. Cotler and F. Wilczek, Bell Tests for Histories. arXiv: 1503.06458 (2015).

\bibitem{Paz}
J. P. Paz and G. Mahler, Proposed test for temporal Bell inequalities. Phys. Rev. Lett. {\bf 71}, 3235 (1993).

\bibitem{Brukner}
Caslav Brukner, Samuel Taylor, Sancho Cheung, and Vlatko Vedral, Quantum Entanglement in Time. arXiv:quant-ph/0402127.

\bibitem{Fritz}
T. Fritz, Quantum correlations in the temporal CHSH scenario. New Journal of Physics {\bf 12}, 083055 (2010).

\bibitem{Emary}
Clive Emary, Neill Lambert, and Franco Nori, Leggett-Garg inequalities. Reports on Progress in Physics {\bf 77},  016001 (2014).



\bibitem{Waldherr}
G. Waldherr et al. Violation of a Temporal Bell Inequality for Single Spins in a Diamond Defect Center. Phys. Rev. Lett. {\bf 107}, 090401 (2011).


\bibitem{GHZ}
D. M. Greenberger, M. A. Horne, A. Shimony and A. Zeilinger, Bell's theorem without inequalities. Am. J. Phys. {\bf 58}(12), 1131 (1990).
\bibitem{Mermin}
D. Mermin, Quantum mysteries revisited. Am. J. Phys. {\bf 58}(8), 731-734 (1990).


\bibitem{Yin}
J. Cotler, L. M. Duan, P. Y. Hou, F. Wilczek, D. Xu, Z.Q. Yin and C. Zu, Experimental Test of Entangled Histories. arXiv:1601.02943v2 (2016).





\bibitem{Tang}
Weidong Tang, Sixia Yu, and C. H. Oh, Greenberger-Horne-Zeilinger Paradoxes from Qudit Graph States. Phys. Rev. Lett. {\bf 110}, 100403 (2013).

\bibitem{Cotler}
J. Cotler and F. Wilczek, Temporal Observables and Entangled Histories. arXiv:1702.05838 (2017).

\bibitem{Zhang}
Chao Zhang, Yun-Feng Huang, Zhao Wang, Bi-Heng Liu, Chuan-Feng Li, and Guang-Can Guo, Experimental Greenberger-Horne-Zeilinger-Type Six-Photon Quantum Nonlocality. Phys. Rev. Lett. {\bf 115}, 260402 (2015).

\bibitem{Su2017}
Zu-En Su, Wei-Dong Tang, Dian Wu, Xin-Dong Cai, Tao Yang, Li Li, Nai-Le Liu, Chao-Yang Lu, Marek \^{Z}ukowski, and Jian-Wei Pan, Experimental test of the irreducible four-qubit Greenberger-Horne-Zeilinger paradox. Phys. Rev. A 95, 030103(R) (2017).




\bibitem{Cerf}
Nicholas J. Cerf, Serge Massar, and Stefano Pironio, Greenberger-Horne-Zeilinger Paradoxes for Many Qudits. Phys. Rev. Lett. {\bf 89}, 080402 (2002).

\bibitem{Ryu1}
Junghee Ryu, Changhyoup Lee, Marek \^{Z}ukowski, and Jinhyoung Lee, Greenberger-Horne-Zeilinger theorem for N qudits. Phys. Rev. A {\bf 88}, 042101 (2013).
\bibitem{Ryu2}
Junghee Ryu, Changhyoup Lee, Zhi Yin, Ramij Rahaman, Dimitris G. Angelakis, Jinhyoung Lee, and Marek \^{Z}ukowski, Multisetting Greenberger-Horne-Zeilinger theorem. Phys. Rev. A {\bf 89}, 024103 (2014).
\bibitem{Lawrence}
Jay Lawrence, Rotational covariance and Greenberger-Horne-Zeilinger theorems for three or more particles of any dimension. Phys. Rev. A {\bf 89}, 012105 (2014).


\bibitem{Jin}
 F Z Jin,  H W Chen,  X Rong, et al., Experimental simulation of the Unruh effect on an NMR quantum simulator. Science China Physics, Mechanics \& Astronomy, {\bf 59}, 630302 (2016).

\bibitem{An2015} Shuoming An, Jing-Ning Zhang , Mark Um, Dingshun Lv, Yao Lu , Junhua Zhang, Zhang-qi Yin, H. T. Quan, and Kihwan Kim, Experimental Test of Quantum Jarzynski Equality with a Trapped Ion System, Nature Physics {\bf 11}, 193 (2015).

\bibitem{Sil}
 I. Silveiro, J M P Ortega,  F J G. De Abajo, Quantum nonlocal effects in individual and interacting graphene nanoribbons. Light: Science and Applications, {\bf 4}, e241 (2015).

\bibitem{Yin2013}
Zhang-qi Yin, Tongcang Li, Xiang, Zhang, L. M. Duan, Large quantum superpositions of a levitated nanodiamond through spin-optomechanical coupling, Phys. Rev. A {\bf 88}, 033614 (2013).

\bibitem{Zych}
Magdalena Zych, Fabio Costa, Igor Pikovski, Caslav Brukner, Bell's Theorem for Temporal Order, arXiv:1708.00248.


\bibitem{Li}
 T Li,  Z Q Yin, Quantum superposition, entanglement, and state teleportation of a microorganism on an electromechanical oscillator. Science Bulletin {\bf 61}, 163-171 (2016).

\bibitem{Wang}
 Z Wang,  C Zhang, Y F Huang, et al., Experimental verification of genuine multipartite entanglement without shared reference frames. Science Bulletin, {\bf 61}, 714-719 (2016).




\end{thebibliography}
\end{document}